# Scattering effect in proton beam windows at spallation targets[*]

Meng Cai， Tang Jingyu， Jing Hantao

(*Institute of High Energy Physics*，*Chinese Academy of Sciences*，*Beijing* 100049，*China*)

**Abstract**： The proton beam window(PBW) is a boundary wall between the high vacuum area in the proton beam line and the helium atmosphere in the helium vessel at a high beam power target. The thermal and mechanical problems of PBW have been studied in foreign spallation neutron sources；however，the scattering effect in PBW is seldom reported which poses serious problems to the target design if not well treated. This paper will report the simulation studies of the scattering effect in PBW. Different materials and different structures of PBW are discussed. Taking China Spallation Neutron Source(CSNS) as an example，a thin single-layered aluminum alloy PBW with edge cooling has been chosen for CSNS-Ⅰ and CSNS-Ⅱ，and a sandwiched aluminum alloy PBW has been chosen for CSNS-Ⅲ. The simulations results of the scattering effect in the presence of beam uniformization at the target by using non-linear magnets at the different CSNS PBWs are presented. The simulations show that the scattering effect at PBW is very important in the beam loss and the beam distribution at the target.

**Key words**： PBW； CSNS； aluminum alloy； Inconel 718； scattering effect； PBW structure

**CLC number**： TL503.4   **Document code**： A   **doi**：10.3788/HPLPB20112310.2773

At spallation neutrons sources，the proton beam window (PBW) is a boundary wall between the high vacuum area in the proton beam line and the helium atmosphere in a helium vessel which contains the target and moderators. The mechanisms of interaction between the proton beam and the PBW include multiple scattering，ionization and nuclear reaction. Multiple scattering is the most important effect leading to beam quality deterioration at the target，and thus a lighter or low-$Z$ material is favored against high-$Z$ material. Energy loss due to ionization is also larger for high-$Z$ material，which mainly has effect on the PBW itself instead of the beam.

Energy deposition in PBW poses serious problems on cooling and mechanical strength against pressure. In foreign spallation neutron sources，the thermal and mechanical problems have been studied to design suitable structures for the PBW；however，the scattering effect at PBW is seldom reported. This paper will report the simulation studies of the scattering effect at PBW，and takes China Spallation Neutron Source(CSNS)[1] as an example.

CSNS is a large facility for multi-disciplinary research using neutron scattering techniques. The project will be constructed in phases. The proton beam power is 100 kW at CSNS-Ⅰ，and will be progressively upgraded to 200 kW at CSNS-Ⅱ and 500 kW at CSNS-Ⅲ. The PBW is in the interface area between the beam transport line[2] and the spallation target[3-4]. The studies on scattering effect are combined with energy deposit and cooling for designing the PBW itself and the target.

## 1　Scattering effect in PBW and simulation method

The scattering effect studies of proton beam passing through PBW have been carried out by using the multipurpose transport Monte Carlo code FLUKA[5] for calculations of particle transport and interactions with matter. As for the beam interaction with material, one can investigate energy loss due to ionization, Coulomb scattering and nuclear reaction and multiple scattering. The small energy loss due to ionization can be ignored here as there is no beam transport element between PBW and target; Coulomb scattering and nuclear reaction will result in immediate beam loss around the PBW and in the way to the target; multiple scattering is the most important effect leading to the emittance blow-up that will change the beam distribution at the target. Fig. 1 is the schematic diagram for simulations of scattering effect at PBW. In the simulations some simplifications have been made, e. g. , the PBW is taken as flat rather than curved and the environment after the PBW is taken as vacuum rather than helium atmosphere. These simplifications do not affect the conclusions. The starting point for this study is at a transport element upstream of the PBW, where the initial beam distribution is given from the beam transport line design. The beam transport down to the target is simulated with FLUKA. In the simulations, the thickness ($t$) of PBW, the distance ($L$) between PBW and target and the PBW structure are variable.

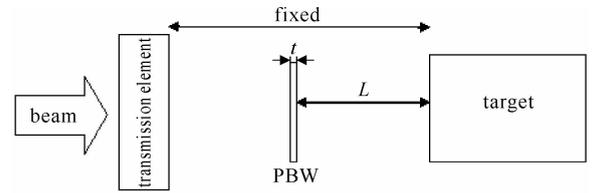

Fig. 1　Schematic diagram for simulations of scattering effect at PBW

## 2　Material and structure of PBW

Four factors, which are closely related to the PBW material and structure, are considered in designing a PBW: energy deposition and cooling, mechanical strength, beam scattering effect and radiation damage. In order to reduce the beam scattering effect, a thin, single-layered and low-$Z$ material PBW is favored. However, the requirements of mechanical strength and cooling prefer thicker and sandwiched design. A compromise has to be made. Until now, three different kinds of PBWs have been used or designed at major spallation neutron sources: sandwiched Inconel window at ISIS and SNS[6], sandwiched aluminum alloy window at J-PARC[7], and multiple aluminum pipes at ESS[8]. SNS has even sought the possibility to use a plasma beam window[9]. Table 1 shows the comparison of PBWs in three spallation neutron sources.

Table 1　Comparison of PBWs at international spallation neutron sources

| | structure | $L$/m | beam energy/ GeV | beam power/ MW | beam loss | PBW material | peak current density/ ($\mu$A·cm$^{-2}$) | footprint |
|---|---|---|---|---|---|---|---|---|
| SNS | Inconel-2 mm, H$_2$O-1.4 mm, Inconel-2 mm, | 2.3 | 1 | 1 | 4% | Inconel 718 | 25 (target) | 200 mm× 70 mm |
| J-PARC | Al-2.5 mm, H$_2$O-3 mm, Al-2.5 mm | 1.8 | 3 | 1 | — | aluminum alloy (A5083-O) | 10 (PBW) | 180 mm× 70 mm |
| ESS | pipe (outer diameter of 6 mm, wall thickness 0.3 mm) | 1.2 | 1.334 | 5 | — | aluminum alloy | 80/150 (target) | 200 mm× 60 mm |

Aluminum alloy and Inconel 718 are common materials of PBW. Stainless steel 316L is also a choice for high power accelerators[10]. The characteristics of an ideal PBW material are high thermal conductivity, high tensile strength at high temperature, low density, high melting temperature, good radiation resistance and excellent weld ability. Table 2 shows the characteristics of three different PBW materials[11]. Aluminum alloy has high thermal conductivity, low density, excellent weld ability and corrosion resistance, but its melting temperature is low. Considering its high thermal conductivity, aluminum alloy is recommended for hun-dred kilowatts accelerators and even megawatts accelerators with special structure. Inconel 718 has good corrosion resistant and good mechanical properties at elevated temperatures (700 ℃) but high density and low thermal conductivity. There is a large beam loss due to the Inconel PBW at the SNS target, as shown in Table 1. Stainless steel 316L has higher thermal conductivity compared with Inconel 718 and is used for helium vessels generally.

Table 2　Characteristics of three different PBW materials

| | density/ (g·mL$^{-1}$) | thermal conductivity/ (W·m$^{-1}$·K$^{-1}$) | linear coefficient of thermal expansion/ ($\mu$m·m$^{-1}$·℃$^{-1}$) | specific heat capacity/ (J·g$^{-1}$·℃$^{-1}$) |
|---|---|---|---|---|
| aluminum alloy(A5083-O) | 2.66 | 117.0 | 16.0 | 0.900 |
| Inconel 718 | 8.19 | 11.4 | 13.0 | 0.435 |
| stainless steel 316L | 7.99 | 21.4 | 19.9 | 0.500 |

| | tensile strength (yield)/MPa | tensile strength (ultimate)/MPa | modulus of elasticity/GPa |
|---|---|---|---|
| aluminum alloy(A5083-O) | 145 | 290 | 70.3 |
| Inconel 718 | 980 | 1 100 | 204.9 |
| stainless steel 316L | 290 | 558 | 193.0 |

For different beam power and PBW materials, three types of structures can be considered: single-layered structure with edge water cooling, double-layered structure with gap water cooling (sandwiched), and multi-pipe structure with inner water cooling, as seen in Fig. 2. PBWs are curved or concavo-convex to have good mechanical properties. Single-layered PBW with edge water cooling and small thickness has the advantages of lower beam loss and beam quality deterioration, but it cannot bear high energy deposition, thus it can be used only in the cases of hundreds of kilowatts or lower. Sandwiched PBW can be used in the cases of about 1 MW or lower. Aluminum alloy thin-pipes structure is a very good design especially in the cases of very high beam power, e.g. 5 MW or higher, but the processing is more difficult.

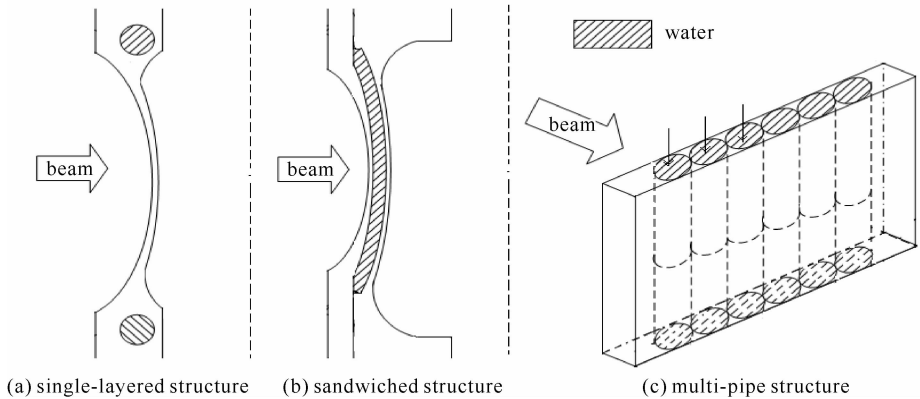

(a) single-layered structure　　(b) sandwiched structure　　(c) multi-pipe structure

Fig. 2　Cross-section schematics of different PBW structures

Considering the beam characteristics of the CSNS, a PBW of single-layer aluminum alloy with edge cooling is perfectly suitable for CSNS-Ⅰ (100 kW) and CSNS-Ⅱ (200 kW). However, at CSNS-Ⅲ (500 kW), a sandwiched PBW with aluminum alloy should be adopted.

## 3　Simulation results with CSNS PBWs

　　The target is the place where the proton beam is converted into neutron flux via spallation processes. Table 3 shows some design features of the CSNS target. Because the beam profile at CSNS-Ⅱ is the same as at CSNS-Ⅰ, we just need to study the scattering effect of CSNS-Ⅰ and CSNS-Ⅲ. Step-like field magnets are used to obtain uniform-like beam distribution at the CSNS target[12], thus it is important to check if the scattering effect affects the distribution uniformization at the target.

Table 3  Design requirements at CSNS target

| | energy/GeV | power/kW | beam current/$\mu$A | footprint at target |
|---|---|---|---|---|
| CSNS-Ⅰ | 1.6 | 100 | 62.5 | 12 cm×4 cm |
| CSNS-Ⅱ | 1.6 | 200 | 125.0 | 12 cm×4 cm |
| CSNS-Ⅲ | 1.6 | 500 | 312.5 | 16 cm×6 cm |

| | peak current density/ ($\mu$A·cm$^{-2}$) | beam proportion within footprint/% | beam loss outside target/W | target size |
|---|---|---|---|---|
| CSNS-Ⅰ | <3.80 | >97.5 | — | 17 cm×7 cm |
| CSNS-Ⅱ | <7.50 | >97.5 | — | 17 cm×7 cm |
| CSNS-Ⅲ | <9.78 | >97.5 | <500 | 17 cm×7 cm |

### 3.1 Simulation results of scattering effect at CSNS-Ⅰ

The simulations show that, the thickness of the PBW should be as small as possible to reduce the scattering effect and the beam loss, and the distance between the PBW and the target should be small to reduce the influence of the scattering effect on the beam distribution at the target. For the former, the PBW of single layer aluminum alloy with edge cooling has been designed, and the thermal and mechanical studies show that the PBW can work reliably at the beam power of 100 kW. For the latter, a compromise between the scattering effect and the overall design of the target assembly has been made to place the PBW at 1.8 m from the target.

Table 4 and Fig. 3 show the simulation results for different thickness of the PBW. Table 5 and Fig. 4 show the simulation results for different distances between the PBW and the target. As having been analyzed qualitatively before, the beam quality at the target decreases and the beam loss increases with the increasing PBW thickness and the increasing distance between PBW and target. As a compromise with the mechanical property, the thickness of 1.0 to 1.5 mm is a good choice. Similarly the distance between the PBW and the target of 1.8 m is also acceptable as a compromise with the target assembly. It should be pointed out here,

Table 4  Simulation results of scattering effect for different PBW thickness at CSNS-Ⅰ ($L=1.8$ m)

| $t$/mm | $\varepsilon_{x\text{-rms}}$/ ($\pi$·mm·mrad) | $\varepsilon_{y\text{-rms}}$/ ($\pi$·mm·mrad) | $\sigma_{x\text{-rms}}$/mm | $\sigma_{x'\text{-rms}}$/mrad | $\sigma_{y\text{-rms}}$/mm | $\sigma_{y'\text{-rms}}$/mrad |
|---|---|---|---|---|---|---|
| 0 | 9.41 | 8.66 | 26.37 | 3.23 | 8.79 | 3.07 |
| 0.5 | 16.96 | 8.91 | 26.39 | 3.30 | 8.83 | 3.11 |
| 1.0 | 21.37 | 9.13 | 26.41 | 3.35 | 8.88 | 3.15 |
| 1.5 | 23.68 | 9.38 | 26.44 | 3.40 | 8.93 | 3.20 |
| 2.0 | 28.77 | 9.56 | 26.47 | 3.48 | 8.97 | 3.24 |
| 2.5 | 30.31 | 9.78 | 26.47 | 3.51 | 9.01 | 3.27 |
| 3.0 | 33.07 | 10.05 | 26.49 | 3.56 | 9.08 | 3.33 |
| 3.5 | 35.51 | 10.19 | 26.52 | 3.62 | 9.10 | 3.35 |
| 4.0 | 37.25 | 10.43 | 26.53 | 3.66 | 9.17 | 3.40 |

| halo(>±2$\sigma_{\text{rms}}$)/% | halo(>±3$\sigma_{\text{rms}}$)/% | $J_{\max}^{(a)}$/ ($\mu$A·cm$^{-2}$) | beam proportion within footprint/% | beam proportion within target/% | beam loss outside target/W | PBW energy deposition/W |
|---|---|---|---|---|---|---|
| 2.25 | 0.03 | 2.63 | 99.62 | 99.995 | 5.0 | 0 |
| 2.62 | 0.21 | 2.60 | 99.37 | 99.83 | 174.1 | 14.8 |
| 3.03 | 0.43 | 2.61 | 99.10 | 99.62 | 385.2 | 31.2 |
| 3.43 | 0.66 | 2.59 | 98.80 | 99.41 | 591.2 | 48.3 |
| 3.80 | 0.79 | 2.62 | 98.56 | 99.28 | 719.3 | 66.0 |
| 4.04 | 0.90 | 2.59 | 98.40 | 99.18 | 816.3 | 81.1 |
| 4.61 | 1.21 | 2.57 | 98.07 | 98.91 | 108 8 | 100.7 |
| 4.82 | 1.32 | 2.58 | 97.87 | 98.80 | 1 203 | 118.4 |
| 5.34 | 1.54 | 2.60 | 97.54 | 98.60 | 1 398 | 136.2 |

a: the peak current density calculated in unit area of 0.48 cm$^2$.

from the simulation results, all the requirements by the target design as given in Table 3 can be met even for the PBW thickness of 4 mm. Fig. 5 shows the beam distribution in phase spaces at the target with the PBW of 1 mm in thickness. One can find that the uniformized distribution within the footprint by using step-like field magnets can be well maintained when the scattering effect at PBW is included, although a large halo is inevitable.

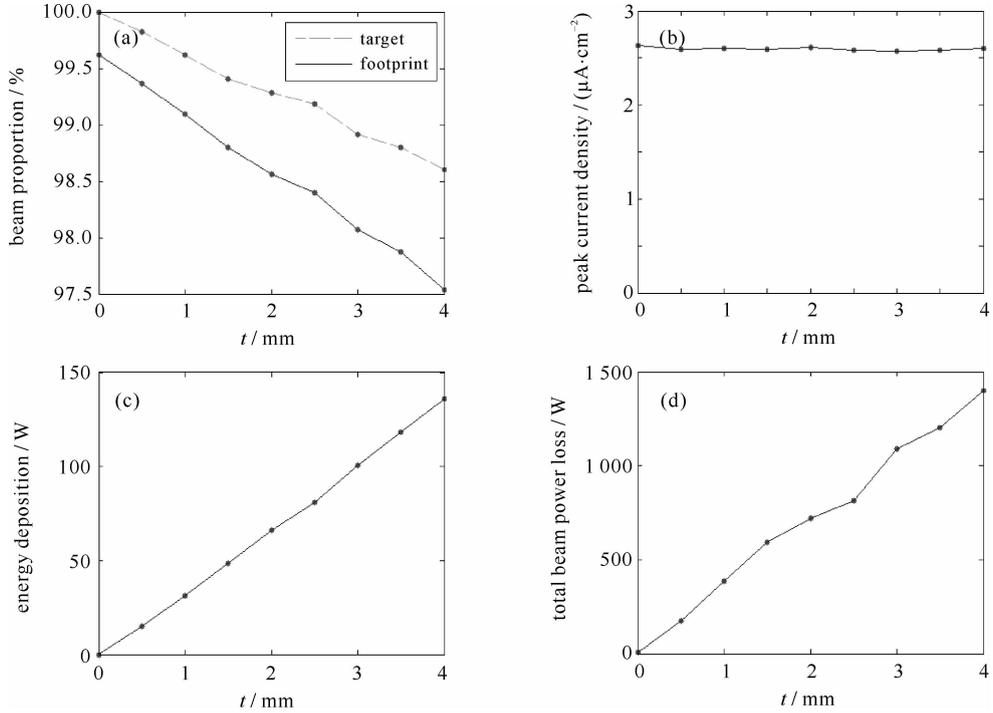

Fig. 3　Simulation results of scattering effect for different PBW thickness at CSNS-I

Table 5　Simulation results of scattering effect for different distances between PBW and target at CSNS-I ($t=1$ mm)

| $L$/m | $\varepsilon_{x\text{-rms}}$/ ($\pi \cdot$ mm $\cdot$ mrad) | $\varepsilon_{y\text{-rms}}$/ ($\pi \cdot$ mm $\cdot$ mrad) | $\sigma_{x\text{-rms}}$/ mm | $\sigma_{x'\text{-rms}}$/ mrad | $\sigma_{y\text{-rms}}$/ mm | $\sigma_{y'\text{-rms}}$/ mrad |
|---|---|---|---|---|---|---|
| 0 | 9.41 | 8.66 | 26.37 | 3.23 | 8.79 | 3.07 |
| 0.5 | 39.70 | 13.15 | 26.38 | 3.58 | 8.81 | 3.35 |
| 1.0 | 26.60 | 10.20 | 26.40 | 3.40 | 8.83 | 3.20 |
| 1.5 | 22.41 | 9.40 | 26.41 | 3.36 | 8.86 | 3.17 |
| 1.8 | 21.37 | 9.13 | 26.41 | 3.35 | 8.88 | 3.15 |
| 2.0 | 19.28 | 9.02 | 26.43 | 3.34 | 8.89 | 3.15 |
| 2.5 | 17.07 | 8.89 | 26.44 | 3.32 | 8.95 | 3.15 |
| halo($>\pm 2\sigma_{\text{rms}}$)/ % | halo($>\pm 3\sigma_{\text{rms}}$)/ % | $J_{\max}^{(a)}$/ ($\mu$A $\cdot$ cm$^{-2}$) | beam proportion in footprint/% | beam proportion in target/% | beam loss outside target/W |
| 2.25 | 0.03 | 2.63 | 99.62 | 100.0 | 5.0 |
| 2.61 | 0.35 | 2.57 | 99.28 | 99.69 | 308.1 |
| 2.77 | 0.40 | 2.57 | 99.19 | 99.65 | 353.1 |
| 2.99 | 0.43 | 2.60 | 99.12 | 99.62 | 381.1 |
| 3.03 | 0.43 | 2.61 | 99.10 | 99.61 | 385.1 |
| 3.17 | 0.45 | 2.57 | 99.07 | 99.60 | 398.2 |
| 3.47 | 0.49 | 2.59 | 98.90 | 99.57 | 430.2 |

a：the peak current density calculated in unit area of 0.48 cm$^2$.

### 3.2　Simulation results of scattering effect at CSNS-Ⅲ

At CSNS-Ⅲ, we plan to use a larger footprint than at CSNS-I to reduce the peak current at the target. Different PBW structures with aluminium alloy have been studied. It is found that single layer structured

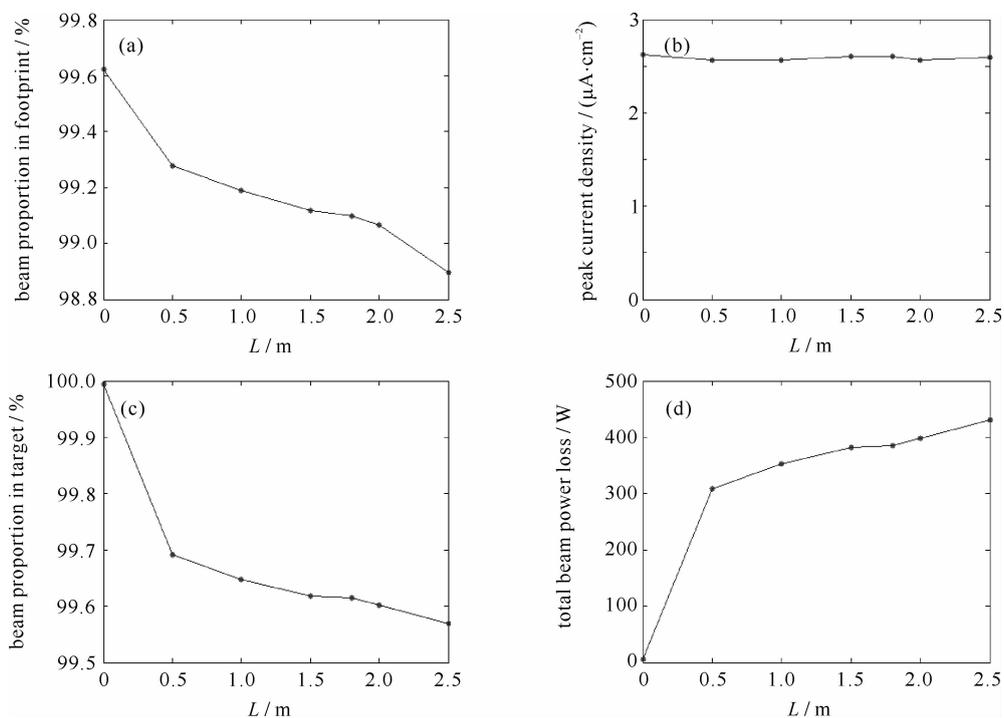

Fig. 4　Simulation results of scattering effect for different distances between PBW and target at CSNS-Ⅰ

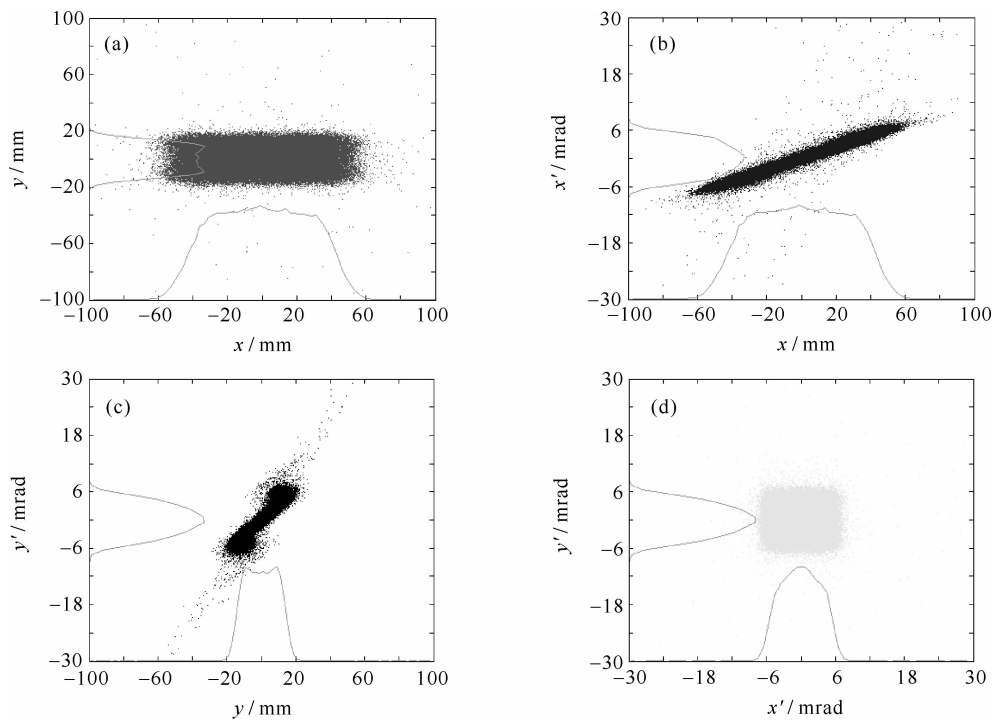

Fig. 5　Phase space plots of beam at target of CSNS-Ⅰ with scattering effect at PBW included ($t=1$ mm, $L=1.8$ m)

PBW can not meet the mechanical requirement due to higher temperature. The sandwiched structure has been adopted. A design scheme of two aluminum alloy plates of 1 mm in thickness and the water gap of 1.5 mm in thickness has been used for the scattering effect studies.

　　Table 6 and Fig. 6 show the simulation results of the scattering effect at PBW at CSNS-Ⅲ. The peak current density at the target can meet the design requirement, but the beam loss outside target is far greater than the requirement. Most beam loss is in the vertical direction because the margin between the vertical footprint size and the target dimension is small. There are two methods to resolve the problem: easing the beam loss requirement and reducing vertical footprint size. Considering that the peak current density at the target becomes critical for the lifetime of the target at CSNS-Ⅲ, it is being considered to ease the requirement for the beam loss outside the target. The beam loss rate at SNS is far larger, where the beam loss has important damage effects on the helium vessel, the target vessel and the moderators and also increases the heat load to the moderators. Similar to the situation at CSNS-I, the uniformized distribution within the footprint can be well preserved.

Table 6    Simulation results of scattering effect at PBW at CSNS-III

| $J_{max}/$ ($\mu A \cdot cm^{-2}$) | beam proportion in footprint/% | beam proportion in target/% | beam loss | PBW energy deposition/W | | |
|---|---|---|---|---|---|---|
| | | | | first layer | water | second layer |
| 7.3 | 98.8 | 99.0 | 5 080 W/1.0% | 155 | 113 | 168 |

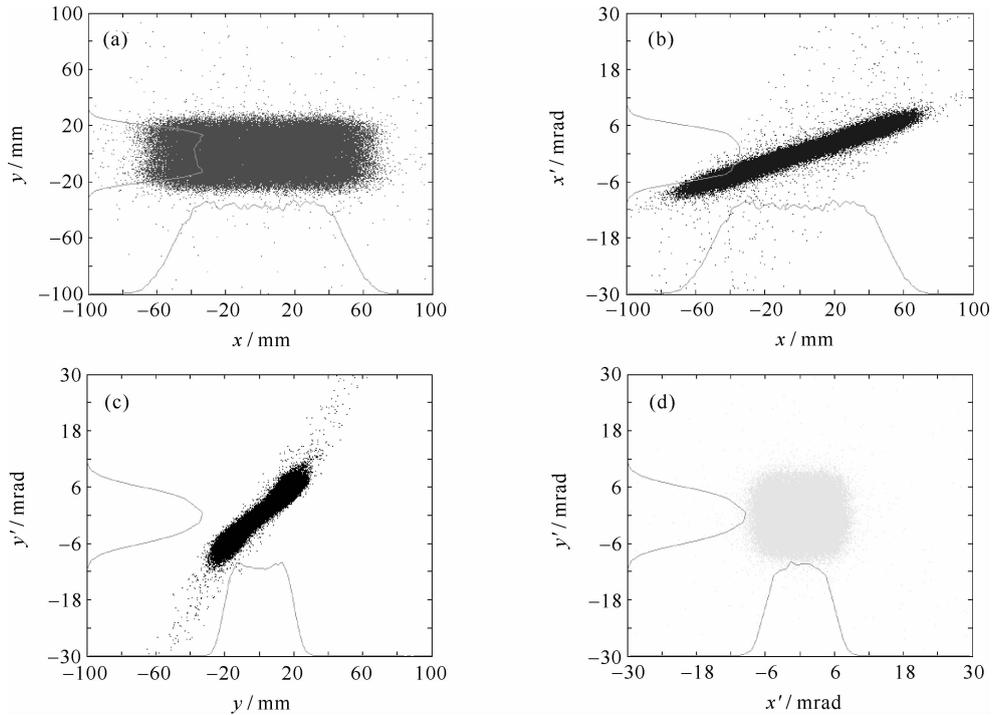

Fig. 6    Phase space plots of beam distribution at target of CSNS-III with scattering effect at PBW

## 4 Conclusions

The structure and material of the PBW at CSNS have been studied. The scattering effect at the PBW has been simulated with FLUKA. The simulations show that the scattering effect at PBW is very important in the beam loss and the beam distribution at the target. A thin single-layered aluminum alloy PBW with edge cooling has been chosen for CSNS-I and CSNS-II, and a sandwiched aluminum alloy PBW has been chosen for CSNS-III. For the former case, the requirements for beam distribution and beam loss at the target can be easily met, however, for the latter case the initial requirements need to be revised as the simulations show that the beam loss is relatively large. The distance between the PBW and the target, an important parameter influencing the scattering effect, has been chosen to be 1.8 m as a compromise with the overall mechanical design of the target assembly. The simulations also show that the uniformized distribution within the footprint at the target can been well preserved when the scattering effect is included.

The studies on scattering effect in PBW provide not only useful information for the interface design between the proton beam transport line and the target at CSNS, but also a good reference for other spallation neutron sources and other high power hadrons beam applications such as accelerator-driven sub-critical systems.

**Acknowledgement**: *The authors would like to thank other CSNS colleagues for the discussions.*

# 散裂靶中质子束窗的散射效应

孟　才，　唐靖宇，　敬罕涛

（中国科学院 高能物理研究所，北京 100049）

**摘　要：**　质子束窗是在高功率靶区中的一个分界窗，它将质子输运线上高真空区域和氦容器中的氦环境分开。在其他散裂中子源中质子束窗的热效应以及机械问题都已经被研究过了，但质子束在该窗中散射效应的研究却很少被报导，然而在靶设计中如果没有处理好质子束窗的散射效应会有很大的问题。报导了质子束窗散射效应的模拟计算结果，包括不同质子束窗的材料和结构选择，并以中国散裂中子源（CSNS）为例，介绍了在 CSNS 一期和二期中质子束窗采用周边水冷的铝合金单层结构，CSNS 三期采用中间水冷的铝合金夹层结构。文中给出了不同结构的质子束窗和不同的与靶距离散射效应对靶上经非线性磁铁均匀化的束流分布的影响的模拟计算结果。模拟结果显示质子窗的散射效应对束流损失和靶上的束流分布有重要的影响。

**关键词：**　质子束窗；　中国散裂中子源；　铝合金；　因科镍合金 718；　散射效应；　质子束窗结构